# Leveraging Video Vision Transformer for Alzheimer's Disease Diagnosis from 3D Brain MRI

Taymaz Akan[a,b], Sait Alp[c], , Md. Shenuarin Bhuiyan[d], Elizabeth A. Disbrow[e,b,f,g], Steven A. Conrad[a], John A. Vanchiere[a,h], Christopher G. Kevil[d,i], and Mohammad A. N. Bhuiyan[a,b] [*]

The Alzheimer's Disease Neuroimaging Initiative (ADNI)[1]

[a]Department of Medicine, Louisiana State University Health Sciences Center at Shreveport, Shreveport, LA, USA
[b]Center for Brain Health, Louisiana State University Health Sciences Center at Shreveport, Shreveport, LA, USA
[c] Department of Artificial Intelligence Engineering, Trabzon, 61335, Turkey
[d]Department of Pathology and Translational Pathobiology, Louisiana State University Health Sciences Center at Shreveport, Shreveport, LA, USA
[e]Department of Pharmacology, Toxicology & Neuroscience, Louisiana State University Health Sciences Center at Shreveport, Shreveport, LA, USA
[f]Department of Neurology, Louisiana State University Health Sciences Center at Shreveport, Shreveport, LA, USA
[g]Department of Psychiatry, Louisiana State University Health Sciences Center at Shreveport, Shreveport, LA, USA
[h]Department of Pediatrics, Louisiana State University Health Sciences Center at Shreveport, Shreveport, LA, USA
[i]Department of Molecular and Cellular Physiology, Louisiana State University Health Sciences Center at Shreveport, Shreveport, LA, USA

[*]**Correspondence to:** Mohammad Alfrad Nobel Bhuiyan, PhD, Division of Clinical Informatics, Department of Medicine, Louisiana State University Health Sciences Center at Shreveport, PO Box 33932, Shreveport, LA 71130-3932. Email: Nobel.Bhuiyan@lsuhs.edu

## Abstract

**Purpose**

Alzheimer's disease (AD) is a neurodegenerative disorder affecting millions worldwide, necessitating early and accurate diagnosis for optimal patient management. In recent years, advancements in deep learning have shown remarkable potential in medical image analysis.

**Methods**

In this study, we present 'ViTranZheimer,' an AD diagnosis approach which leverages video vision transformers to analyze 3D brain MRI data. By treating the 3D MRI volumes as videos, we exploit the temporal dependencies between slices to capture intricate structural relationships. The video vision transformer's self-attention mechanisms enable the model to learn long-range dependencies and identify subtle patterns that may indicate AD progression. Our proposed deep learning framework seeks to enhance the accuracy and sensitivity of AD diagnosis, empowering clinicians with a tool for early detection and intervention. We validate the performance of the video vision transformer using the ADNI dataset and conduct comparative analyses with other relevant models.

**Results**

---

[1] Data used in preparation of this article were obtained from the Alzheimer's Disease Neuroimaging Initiative (ADNI) database (adni.loni.usc.edu). As such, the investigators within the ADNI contributed to the design and implementation of ADNI and/or provided data but did not participate in analysis or writing of this article. A complete listing of ADNI investigators can be found at:
http://adni.loni.usc.edu/wp-content/uploads/how_to_apply/ADNI_Acknowledgement_List.pdf

The proposed ViTranZheimer model is compared with two hybrid models, CNN-BiLSTM and ViT-BiLSTM. CNN-BiLSTM is the combination of a convolutional neural network (CNN) and a bidirectional long-short-term memory network (BiLSTM), while ViT-BiLSTM is the combination of a vision transformer (ViT) with BiLSTM. The accuracy levels achieved in the ViTranZheimer, CNN-BiLSTM, and ViT-BiLSTM models are 98.6%, 96.479%, and 97.465%, respectively. ViTranZheimer demonstrated the highest accuracy at 98.6%, outperforming other models in this evaluation metric, indicating its superior performance in this specific evaluation metric.

**Conclusion**

This research advances the understanding of applying deep learning techniques in neuroimaging and Alzheimer's disease research, paving the way for earlier and less invasive clinical diagnosis.

**Keywords:** *Alzheimer's disease, Early diagnosis, Multiclass classification, Deep learning, Vision transformer, ViViT.*

# 1 Introduction

Until now, Alzheimer's disease (AD) has persisted as one of the most debilitating chronic neurological disorders, predominantly impacting individuals aged 65 and above [1]. From a clinical standpoint, AD is marked by a wide range of symptoms, such as memory problems, aphasia, apraxia, agnosia, problems with spatial skills, executive dysfunction, and changes in personality and behavior [2,3]. With the increasing aging population, AD is anticipated to emerge as a significant global burden in the forthcoming decades. According to the World Health Organization (WHO) [4], the prevalence of AD was estimated to be 50 million individuals in 2015, with projections indicating it will triple by the year 2050 [1,2]. Typically, the advancement of AD can be categorized into three stages: early, middle, and late [5]. The accurate diagnosis of AD in its early stage, mild cognitive impairment (MCI), is of utmost importance due to the rapid rise in its prevalence. This is essential for timely treatment and potentially delaying the progression of AD [6]. At the moment, there is no cure for AD, and treatments can only slow the disease's progress [4]. Early diagnosis is very important because the new drugs target early-stage disease [7]. AD diagnosis can be based on various techniques, including cognitive and neuropsychological tests, neurological examinations, biomarkers, genetic testing, and neuroimaging. Structural magnetic resonance imaging (sMRI) can be used to find brain pathology (such as atrophy, tumors, and lesions) and rule out causes of cognitive deficits other than AD. sMRI is a non-invasive process used in clinical settings for suspected AD without injections, surgery, or radiation exposure, making it safe and well-tolerated by patients.

In recent years, advances in deep learning have revolutionized medical image analysis [8]. The use of this technique has demonstrated remarkable potential in various facets of medical imaging, bringing substantial advantages to healthcare and diagnosis. Like other areas of medical image analysis, brain MRI has benefited greatly from the application of deep learning [9,10].

Deep learning techniques in image processing involve the use of artificial neural network architectures, such as Convolutional Neural Networks (CNNs) [11], Recurrent Neural Networks (RNNs) [12–14], and Vision Transformers (ViTs) [15]. CNNs use convolutional layers to scan images with learnable filters, detecting patterns and features at different scales. RNNs handle sequential data but can also be adapted for image processing tasks like image captioning. ViTs use self-attention mechanisms to process images as sequences of tokens, capturing spatial relationships and dependencies among image patches. CNNs are effective in computer vision tasks like image classification [16], object detection/tracking [17,18], and semantic segmentation [19]. They are widely used in medical image analysis, particularly in 2D and 3D ultrasound and MRI images [20].

Since an MR brain scan is 3D data made of stacked 2D slices, a model that can describe the whole brain is needed. A 3DCNN model to maintain inter-slice relationships is the most straightforward approach, as 2DCNN cannot do so and require an additional sequential model like LSTM. However, 3DCNN has many parameters and computation is high. Slice-based methods split 3D neuroimages into 2D slices for MRI analysis [21]. However, slice-based methods analyze 3D neuroimages into 2D slices in MRI scans. When classifying 3D-MRI scans using slide-based methods, spatial features in each slice and their relationships must be considered. Pre-trained 2D-CNNs can extract slice-specific features. Deep sequence modeling models like RNN-based models can then model slice dependency.

In contrast to CNN-based models, ViT-based models lack an inductive bias; therefore, it is necessary to train these models with massive amounts of information. Since ViT models are pre-trained on a large number of two-dimensional data, their use on two-dimensional data becomes efficient. Since the MRI is 3D and the available data are insufficient to train the 3D ViT, it is unreasonable to expect the 3D ViT to produce similarly effective results as the standard ViT. The common method involves splitting 3D data into 2D slice arrays from plane to take advantage of transfer learning with a pre-trained 2D network model. ViT extracts slice features separately, utilizing deep sequence models like RNN to model slice dependency, ensuring relationships between slices are maintained [22–24]. While the use of 3D-ViT in AD applications is still limited, emerging studies show promise in leveraging their capacity to capture complex spatial features in 3D neuroimaging data [25,26].

In this study, we present ViTranZheimer, an approach for AD diagnosis that leverages video vision transformers to analyze 3D brain MRI data. ViTranZheimer treats 3D brain MRI as distinct 2D slices or rely on CNNs with recurrent layers such as LSTMs, allowing it to fully exploit inter-slice dependencies. This differs from existing methods in its use of video vision transformers, which utilize self-attention mechanisms to capture both short- and long-range dependencies across the entire 3D volume.

## 2  Materials and method

In this study, we propose an end-to-end approach for the classification of T1-weighted MRI data using the Video Vision Transformer (ViViT) model, applied directly to the entire MRI voxel. We introduce a novel approach where each slice of MRI is treated as a frame in a video, enabling the use of the Video Vision Transformer (ViViT) model for deep video classification. Our previous method [27] utilized Vision Transformers (ViT) for each 2D slice independently, followed by a sequential classification model to combine the features, but it lacked an end-to-end framework. In that approach, the processing of slices and their subsequent classification were handled separately. This separation limited the ability to optimize the feature extraction and classification processes in a fully integrated, end-to-end manner.

End-to-end models offer several key advantages, particularly in complex tasks like medical imaging. First, they enable joint optimization, where all layers from feature extraction to classification are trained together, leading to better overall performance. This approach reduces the need for manual intervention, as the model automatically learns relevant features, minimizing human error and making the process more efficient. End-to-end models also excel at capturing complex relationships, such as spatio-temporal dependencies in 3D MRI scans, which improves accuracy. Additionally, they simplify the training pipeline by consolidating the entire process into a unified framework, streamlining the learning process and reducing potential errors from separate processing steps [28]. However, our approach leverages ViViT to capture spatio-temporal dependencies across the full 3D structure of the brain. This allows the model to maintain inter-slice relationships and extract more comprehensive features from the entire MRI volume.

### 2.1  Structural MRI data

The ADNI [29,30] dataset, created in 2003, is a comprehensive set of data on AD and related disorders. It includes data from various sources, including cognitive tests, genetic analysis, PET, and MRI. The ADNI MRI Core has created standardized analysis sets for analysis, allowing for more careful comparisons and meaningful comparisons of algorithms. This study used a medium shared standard data collection, ADNI1: Complete 3Yr 3T, to compare different brain structures and algorithms. The dataset included 351 image scans, where were randomly split into 60% training, 20% testing, and 20% validation sets. The study aimed to ensure meaningful comparisons and reduce the possibility of differences in algorithm performance due to different input data. **Table 1** shows details of the data collection.

**Table 1.** The details of data collections

| | # Image Scans | NC | MCI | AD | Male | Female | Age |
|---|---|---|---|---|---|---|---|
| ADNI1: Complete 3Yr 3T | 351 | 129 | 145 | 77 | 194 | 157 | $75 \pm 7.07$ |

## 2.2 MRI pre-processing

The processing and analysis of sMRI data consists of several steps. These steps include aligning the images, registering them to a standard template such as MNI space, segmenting the images into different tissue types such as gray matter, white matter, and cerebrospinal fluid, and conducting voxel-based analyses (See Figure 1 and Table 2). Image registration is crucial for ensuring the spatial alignment of MRI scans, as it helps standardize them with a fixed-size template.

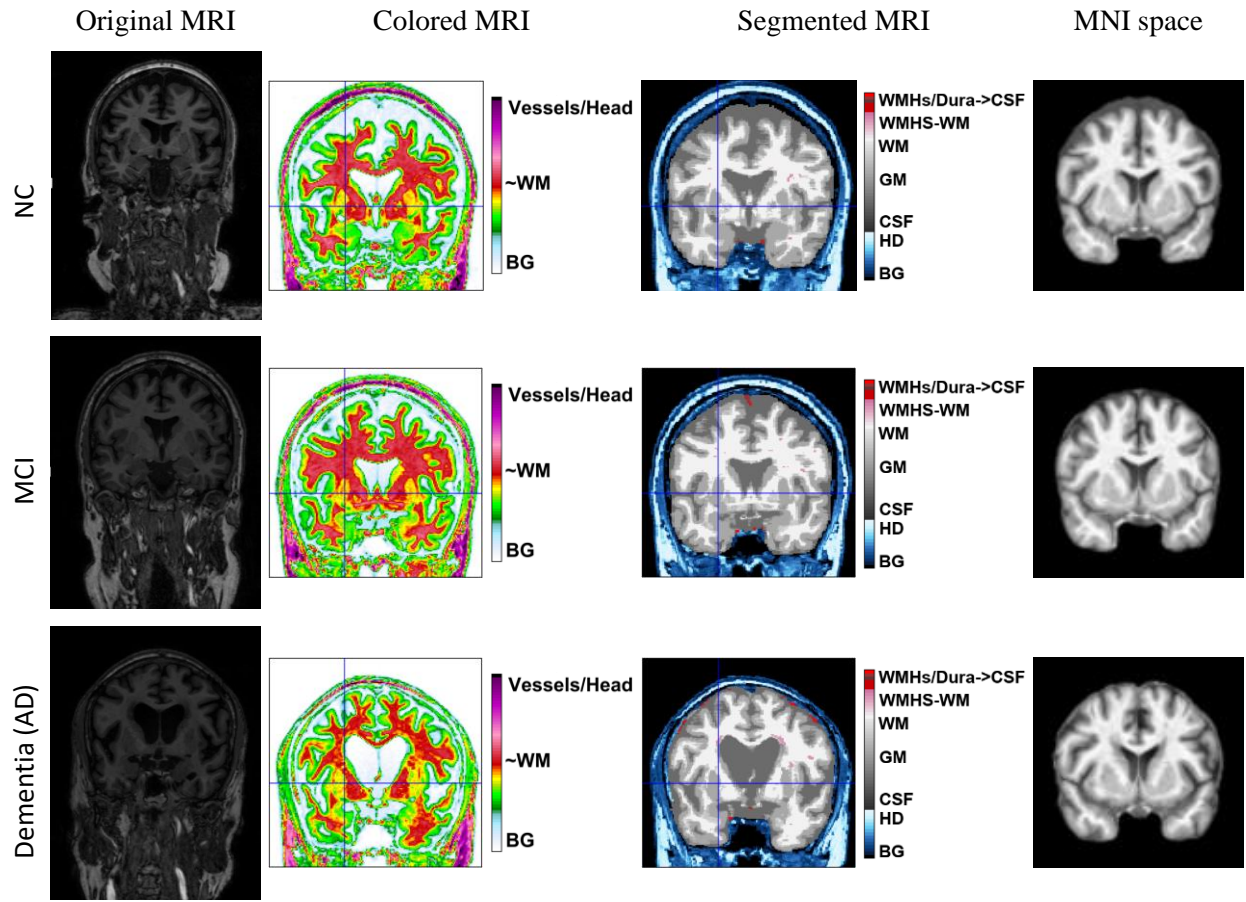

**Figure 1.** A sample of MRI slices from subjects with NC, MCI, and dementia. The first columns show the original MRI, the second, third, and fourth columns show colored MRI, brain tissue segmented MRI, and normalized MRI with template, respectively. The MRI scans are first segmented using CAT12, followed by skull-stripping to remove non-brain tissues. After segmentation and skull-stripping, the images are normalized to MNI space to ensure consistency and alignment across subjects.

The Montreal Neurological Institute's MNI space is a common coordinate system for neuroimaging, based on a template brain from hundreds of healthy MRI scans. This space provides a precise reference slice for analyzing brain activity across various studies and research teams. Skull stripping, normalization, and image registration are performed using Statistical Parametric Mapping 12 (SPM 12) [31] to wrap MRI scans into the MNI-152 space. Moreover, only the central 32 slices containing significant slices are selected. The remaining slices are excluded due to background information unrelated to brain tissue

**Table 2.** The details of MRIs for subjects with NC, MCI, and dementia

|    | **Image processing quality:** |        | **Volumes**     | CSF  | GM   | WM        |
|----|-------------------------------|--------|-----------------|------|------|-----------|
| NC | Resolution                    | 83.47% | Absolute volume | 301  | 593  | 462 $cm^3$ |
|    | Bias                          | 98.01% | Relative volume | 22.2 | 43.7 | 34.1%     |
|    | weighted image quality (IQR)  | 82.46% | TIV             | 1356 $cm^3$ | | |
| MCI| Resolution                    | 83.58% | Absolute volume | 396  | 708  | 617 $cm^3$ |
|    | Bias                          | 98.16% | Relative volume | 23.0 | 41.1 | 35.9%     |
|    | weighted image quality (IQR)  | 83.34% | TIV             | 1721 $cm^3$ | | |
| AD | Resolution                    | 83.51% | Absolute volume | 691  | 699  | 601 $cm^3$ |
|    | Bias                          | 97.13% | Relative volume | 34.7 | 35.1 | 30.2%     |
|    | weighted image quality (IQR)  | 85.08% | TIV             | 1991 $cm^3$ | | |

TIV (Third module: estimate total intracranial volume); CSF (Cerebrospinal fluid); GM (Gray matter); WM (White matter)

## 2.3 End-to-End classification architecture

This section provides a comprehensive presentation of the proposed framework. We employed 3D-MRI in our approach for the classification of AD. ViTranZheimer involves feeding entire 3D brain scans as input to models, rather than slicing them into 2D images.

By treating the 3D MRI slices as consecutive frames in a video, we leverage a complex video understanding deep-based model to handle the long sequences of slices/frames. In this study, the recently introduced ViViT [32], a transformer-based video encoder, is employed for AD classification tasks. We used spatio-temporal attention as it lets a model capture both spatial features (information each individual frame) and temporal dynamics (changes between frames). It leads the model to discover complex patterns and connections in the MRI that cover time and space.

For a given 3D MRI volume with dimensions $T \times H \times W$ (where $T$ is the number of slices, and $H$ and $W$ are the height and width of each slice), we divide the volume into tubelets (see Figure 2). Each tubelet spans a small region both spatially and temporally (across adjacent slices). If the tubelet size is $P_t \times P_h \times P_w$, the number of tokens in each dimension is calculated as: $n_t = \left\lfloor \frac{T}{P_t} \right\rfloor, n_h = \left\lfloor \frac{H}{P_h} \right\rfloor, n_w = \left\lfloor \frac{W}{P_w} \right\rfloor$.

Then, the volumes are flattened in order to create video tokens. These tokens are extracted from the temporal, height, and width dimensions of a tubelet with dimensions $t \times h \times w$. This method incorporates spatio-temporal information into the tokenization process in an intuitive manner. The tubelets are flattened and projected into an embedding space as follows:

$$z_0 = [E(\text{Tubelet}_1), E(\text{Tubelet}_2), \ldots, E(\text{Tubelet}_N)] + E_{\text{pos}} \quad (1)$$

Here, $E(\cdot)$ is a learnable linear embedding function, and $E_{\text{pos}}$ adds positional encodings that indicate the position of each tubelet in the original 3D volume.

Once the tubelets are embedded, the 3D MRI data is fed into the Transformer architecture. The core innovation here is the Spatio-temporal attention mechanism, which allows the model to process both spatial and temporal information from the MRI scans simultaneously.

Each embedded token represents a 3D tubelet, containing spatial and temporal information. The transformer uses the self-attention mechanism to compute the relationships between these tokens. For each token, queries (Q), keys ( K ), and values (V) are computed, and the attention scores $\alpha$ are determined by the dot-product attention formula:

$$\alpha_{ij} = \frac{\exp\left(Q_i K_j^T / \sqrt{d_k}\right)}{\sum_{j=1}^{N} \exp\left(Q_i K_j^T / \sqrt{d_k}\right)} \quad (2)$$

The self-attention layer allows the model to attend to relevant spatial and temporal regions across the 3D MRI data, ensuring that it can learn long-range dependencies between slices. The attention mechanism computes a weighted sum over the values $V_j$ for each token $i$:

$$\text{Attention}(Q, K, V) = \sum_{j=1}^{N} \alpha_{ij} V_j \quad (3)$$

This is applied across all tubelets, so the model can focus on specific parts of the 3D volume that are more important for Alzheimer's diagnosis.

Before feeding the tubelets into the transformer, we need to convert them into a suitable representation. This is where Patch Embedding comes in, which is conceptually similar to the patch-based embedding used in ViT. For each tubelet, we flatten the 3D patch into a 1D vector per: $\mathbf{x}_i = \text{Flatten}(\text{Tubelet}_i) \in \mathbb{R}^{P^2 \times P_t}$. These vectors are then passed through a linear projection layer: $\mathbf{z}_i = W_E \mathbf{x}_i + b_E$. Where $W_E$ and $b_E$ are the learnable parameters of the projection layer. This gives us a sequence of embeddings for the tubelets, which serve as input to the transformer.

Moreover, ViViT architecture uses multi-head self-attention to capture multiple relationships between the tubelets in different subspaces. For each head, the attention mechanism is applied independently, and the results are concatenated. Mathematically, this can be expressed as: $\text{MHSA}(Q, K, V) = [\text{head}_1, \text{head}_2, \ldots, \text{head}_h] W_O$. Each attention head $\text{head}_i$ is computed as: $\text{head}_i = \text{Attention}(Q_i, K_i, V_i)$.

Where $Q_i, K_i, V_i$ are projections of the input embeddings for the $i$-th attention head. This allows the model to attend to different parts of the 3D MRI volume simultaneously, capturing diverse spatial and temporal features. After applying multi-head self-attention, the output is passed through a FeedForward Network (FFN) for further processing. The FFN consists of two fully connected layers with a ReLU activation: FFN $(x) = $ ReLU $(xW_1 + b_1)W_2 + b_2$. This helps the model to learn more complex transformations of the data before classification.

After passing through multiple layers of spatio-temporal attention and feed-forward networks, we aggregate the information from all tubelets using the CLS token (classification token). The CLS token gathers global information from the entire 3D MRI volume and is used for the final classification task. The final classification is computed per: $\hat{y} = $ softmax $(W_{cls}z_{CLS} + b_{cls})$. Where $W_{cls}$ and $b_{cls}$ are the parameters learned for the classification head, and $z_{CLS}$ is the embedding of the CLS token.

Figure 3 provides a visual summary of the ViTranZheimer. Before treating the MRI slices as consecutive frames in a video, the preprocessing of MRI images is conducted using CAT12 for tasks such as spatial normalization, segmentation, and other structural adjustments. Then, these frames are embedded into tokens and inputted into a transformer encoder, which uses self-attention mechanisms to extract features. The output is passed through a multi-layer perceptron (MLP) head for classification.

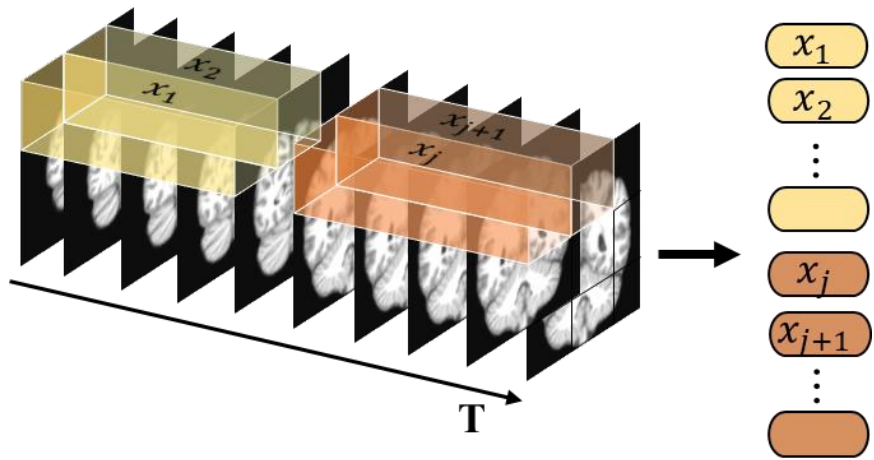

**Figure 2.** Tubelet embedding: It involves the extraction and linear embedding of nonoverlapping tubelets that cover the entire spatio-temporal input volume.

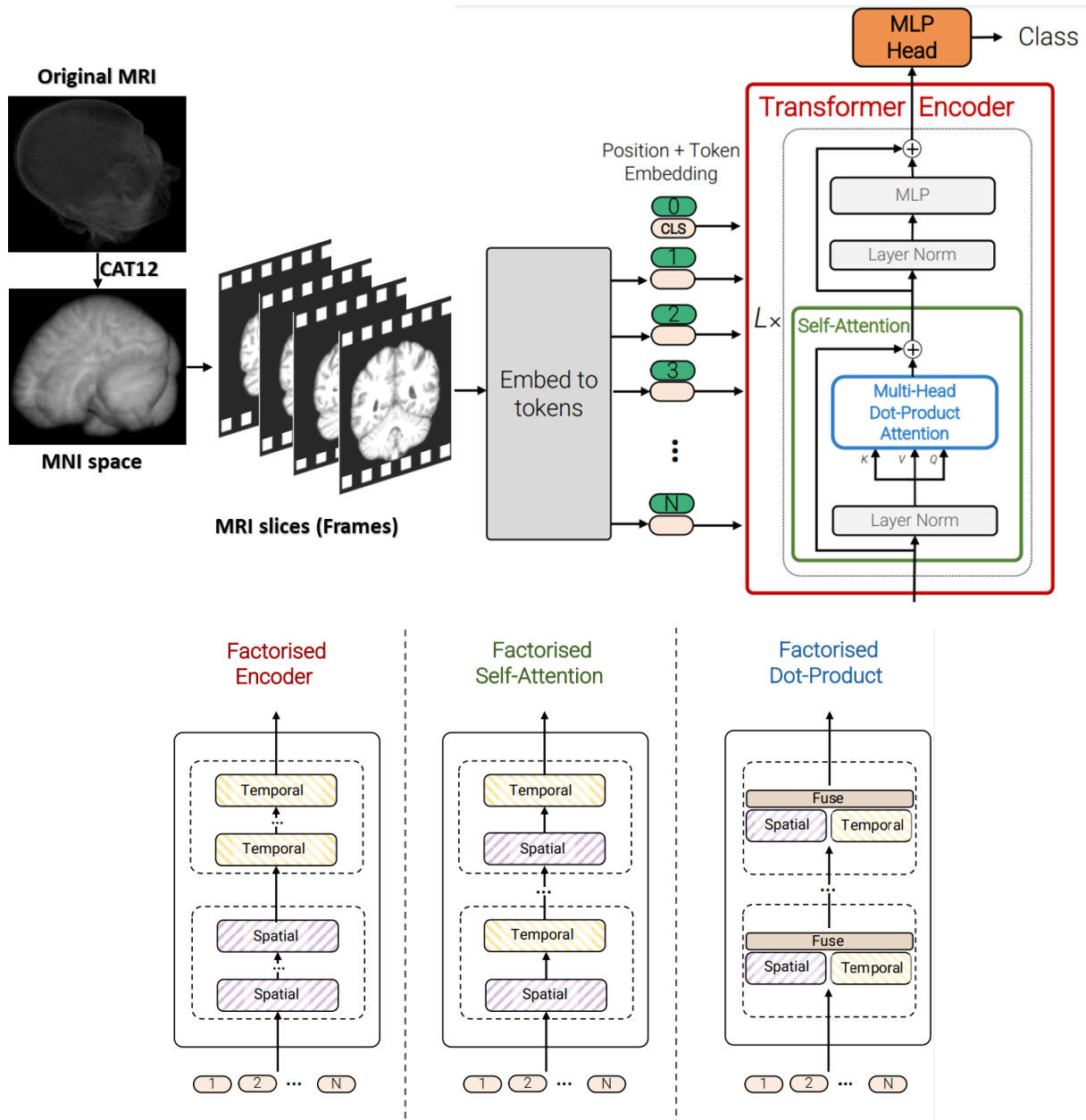

**Figure 3.** The pipeline proposed by ViTranZheimer. CAT12 was utilized to pre-process MRI images (image registration to standardize the images and skull stripping to reduce biases by ensuring consistent voxel intensities). ViViT was applied to whole frames to determine their characteristics and interdependencies.

## 2.4 Setting

The proposed framework is constructed using the Keras framework and the TensorFlow backend. All experiments were conducted on a workstation with an NVIDIA GeForce RTX 4080 GPU and 64 GB of RAM. To test the proposed method, we conducted experiments on multi-classification tasks (CN, MCI, and AD). Table 3 contains a list of the models' specifications. The model was trained from scratch with the associated data using the Adam optimizer for a total of 1500 epochs, with a batch size of 128 and a learning rate of 1e-4. The model has a total of 466,115 trainable parameters. The configuration of training parameters is summarized in Table 3. As the callback, the model checkpoint was configured to save only the optimal solution discovered during training, based on the loss function evaluation performed during validation. During training, the metric on a validation set is monitored, and the model checkpoint is only saved when the metric improves. The model checkpoint is saved whenever a metric improves on a validation set during training.

### 2.5 Performance Evaluation

We used various classification performance metrics to evaluate the performance of the models. These include precision, F1-score, recall, and overall accuracy. Below are the definitions and equations for each of these metrics:

- **Precision** measures the accuracy of positive predictions and is calculated as:

$$\text{Precision} = \frac{\text{True Positives (TP)}}{\text{True Positives (TP)} + \text{False Positives (FP)}} \quad (4)$$

- **Recall** is essential for minimizing false negatives, ensuring that individuals with a condition are correctly identified and receive timely medical attention and is calculated as:

$$\text{Recall} = \frac{\text{True Positives (TP)}}{\text{True Positives (TP)} + \text{False Negatives (FN)}} \quad (5)$$

- **F1-Score** balances precision and recall, providing a useful measure when both false positives and false negatives are critical, as is often the case in medical diagnoses. It is the harmonic mean of precision and recall, providing a balance between the two metrics:

$$\text{F1-Score} = 2 \times \frac{\text{Precision} \times \text{Recall}}{\text{Precision} + \text{Recall}} \quad (6)$$

- **Overall Accuracy** offers a general sense of how often the model makes correct predictions, but in medical applications, precision and recall often carry more weight since the impact of false negatives or false positives can lead to missed diagnoses or overtreatment. It is calculated per:

$$\text{Accuracy} = \frac{\text{True Positives (TP)} + \text{True Negatives (TN)}}{\text{Total Number of Predictions}} = \frac{\text{TP} + \text{TN}}{\text{TP} + \text{TN} + \text{FP} + \text{FN}} \quad (7)$$

**Table 3.** Details of model variants

| Parameter name | Values |
| --- | --- |
| **Hyper parameters of ViViT** | |
| Optimizer | Adam |
| Batch size | 128 |
| Epoch | 1500 |
| Input Shape | (32, 64, 64, 1) |
| Layer norm | 1e-6 |
| Learning rate | 1e-4 |
| Number of heads | 16 |
| Number of Layers | 16 |
| Patch size | (32, 16, 16) |
| Projection dim | 32 |

## 3 Experimental Results and evaluation

### 3.1 Configuration

The classification task was done on images in the coronal plane. We implemented different baseline architectures to make comparisons with the proposed method. We compared the accuracy of our classification (ViTranZheimer) using CNN and Bi-LSTM [33], as well as ViT and Transformer [27]. The CNN and ViT were utilized in order to derive the attributes of the T1-weighted MRI slices, and the Bi-LSTM model was utilized in order to classify the sequential features while maintaining the inter-association between the slices. The details of baseline model variants are listed in Table 4. Afterwards, we compared metrics like classification accuracy, model precision, recall, and F-score.

Repeated 10-fold stratified cross-validation and testing were adopted for the evaluation of ViTranZheimer on the ADNI dataset. The cross-validation and test were done 10 times by moving the start subset of the cross-validation and test settings to the next subset. This way, each sample of all the datasets was only used for the test once. This approach offers several advantages, including the ability to reduce variance in performance estimates, utilize more training data, prevent overfitting, and ensure consistent, fair model evaluation across different machine learning models.

In this study, we also compare our proposed method with state-of-the-art models to fully evaluate its effectiveness. To this end, various models from the literature were compared with the proposed method.

**Table 4.** Details of baseline model variants

| Parameter name | | Values |
|---|---|---|
| **CNN** | | |
| Model | | ResNet-101 |
| Trained with | | ImageNet-1k (contains 1,200,000 images) at resolution 224x224. |
| Params | | 44.7M |
| Depth | | 209 |
| Size of last layer (Feature extractor) | | $1 \times 2048$ |
| **Bi-LSTM** | | |
| Layers | | 6 |
| Hidden Units | | 64 |
| Epoch | | 100 |
| Batch Size | | 25 |
| | learning rate | 1e-4 |
| Adam Optimizer | $\beta_1$ | 0.9 |
| | $\beta_2$ | 0.999 |
| | $\epsilon$ | 1e-07 |
| Loss Function | | Sparse Categorical Cross entropy |
| Dropout | | 0.15 |
| Activation | | tanh |
| Input Dimension | | $32 \times 2048$ |
| Output Dimension | | #classes (2/3) |
| Params | | 538,515 |

## 3.2 Results

Our findings indicate that our model demonstrated the highest level of accuracy in classifying disease status for individuals with NC, MCI, and AD in various clinical diagnosis tasks (see Table 5). Context of multiclass disease classification on the ADNI1: Complete 3Yr 3T dataset, three distinct neural network architectures exhibited impressive performance.

The CNN-Bi-LSTM model achieved an accuracy of 96.479%, with strong precision, recall, and F-score values of 0.96, indicating its proficiency in correctly classifying disease categories. ViT-Bi-LSTM surpassed this performance, achieving an accuracy of 97.465% and maintaining high precision, recall, and F-score values at 0.97. Notably, ViTranZheimer outperformed both models, with an accuracy of 98.6%, accompanied by precision, recall, and F-score metrics also at 0.97.

**Table 5.** Results for multiclass disease classification on ADNI1: Complete 3Yr 3T.

| Architecture | ACC | Precision | Recall | F-score |
|---|---|---|---|---|
| CNN-Bi-LSTM | 96.479% (±2.205) | 0.96 | 0.96 | 0.96 |
| ViT-Bi-LSTM | 97.465% (±2.164) | 0.97 | 0.97 | 0.97 |
| ViTranZheimer | **98.6% (+/-1.4)** | 0.97 | 0.97 | 0.97 |

In Figure 4, we present the confusion matrices depicting the outcomes of our multiclass disease classification analysis.

In the AD class, ViTranZheimer attains a 97% true positive rate, correctly identifying 97% of AD cases, while maintaining a perfect 100% true negative rate for non-AD cases. ViT-Bi-LSTM also performs well for AD, achieving a true positive rate of 93%. CNN-Bi-LSTM, while decent, lags slightly behind with a truly positive rate of 93%. In the case of NC, ViTranZheimer demonstrates perfect sensitivity and specificity, correctly classifying all NC cases without any false positive or false negatives. ViT-Bi-LSTM and CNN-Bi-LSTM both exhibit excellent accuracy for NC and follow ViTranZheimer with true positive rates of 99% and 97%, respectively. Regarding Mild Cognitive Impairment, we obtained a 98% true positive rate and a 99% true negative rate, with only 1% false positives and 1% false negatives. All three methods demonstrate strong performance, with ViTranZheimer and ViT-Bi-LSTM achieving a 98% true positive rate, closely followed by CNN-Bi-LSTM with a slightly higher 99% true positive rate.

Overall, ViTranZheimer excels in differentiating between AD, NC, and MCI, offering a promising diagnostic approach for AD utilizing video vision transformers and 3D brain MRI data.

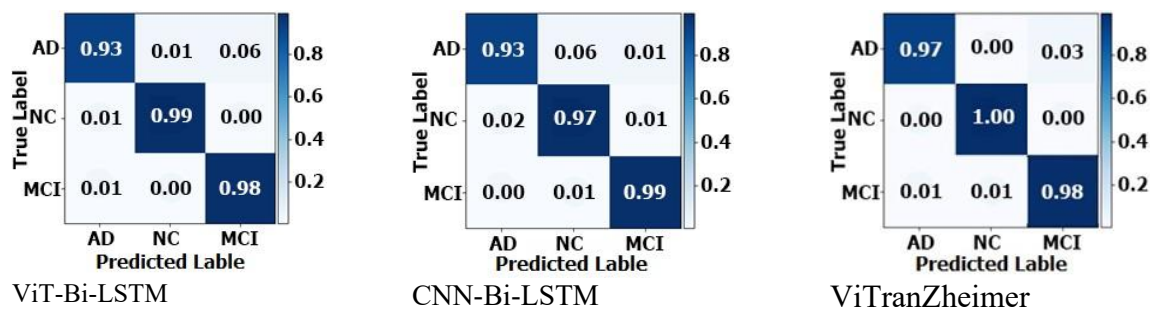

ViT-Bi-LSTM          CNN-Bi-LSTM          ViTranZheimer

**Figure 4.** Confusion matrices for multiclass disease classification (NC, MCI, and AD) on ADNI1: Complete 3Yr 3T.

Additionally, we present a comparative analysis of our findings in relation to existing studies that have been reported in the literature. Table 6 presents the numerical outcomes pertaining to the multi-classification tasks conducted on the ADNI datasets.

**Table 6.** Performance compared with state-of-the-art methods for multiclass disease classification.

| work | Input | Image scans (NC/MCI/AD) | Method | NC/MCI/AD classification (%) | | | | |
|---|---|---|---|---|---|---|---|---|
| | | | | ACC | recall | Precision | SPE | F1-Score |
| [34] | Slice based | 300/300/300 | VGGNet. 16 | 91.85 | - | - | - | - |
| [35] | Voxel based | 70/70/70 | 3D CNN | 89.10 | - | - | - | - |
| [36] | Slice based | 229/243/188 | ResNet-18 | 56.80 | - | - | - | - |
| [37] | Slice based | 169/234/101 | 2D CNN | 96.00 | 96.0 | - | 98.0 | - |
| [38] | Voxel based | 574/808/346 | 3D ResNet | 87.00 | - | - | - | - |
| [39] | Voxel based | 207/215/193 | 3D VGGNet | 91.13 | - | - | - | - |
| [40] | Slice based | 50/50/50 | VGGNet-16 | 95.73 | 96 | 96.33 | - | 95.66 |
| [41] | Voxel based | 351/297/221 | 3D DenseNets | 97.52 | 97 | 97.13 | - | **97.02** |
| [42] | Patch based | 475/224/70 | 3D CNN | 97.48 | 95.33 | 97.33 | | 97.0 |
| [43] | Slice based | 25/13/25 | ResNet18 & DenseNet121 | 98.21 | 98.14 | - | **98.14** | - |
| Our method | Slice-based | 129/145/77 | ViViT | **98.6** | **99.0** | 97.0 | - | 97.0 |

Comparing our study to others was difficult due to the lack of precise data sets and training procedures. ADNI is useful for AD research, but most researchers choose data based on personal preference. Most of them have unequal sample sizes and no data overlap. Additionally, most studies have used binary (CN/AD) classification. The biggest challenge is multi-classifying data samples to distinguish MCI from AD and NC. All Table 5 data were chosen for personal preference and accordance. The third column of these tables shows that not only the number of samples is not equal, but the data may also not overlap. To this end, this study used a medium-sized standard data collection named ADNI1: Complete 3Yr 3T that is publicly available.

## 4 Discussion

MCI is a crucial intermediate stage between normal cognitive aging and more serious conditions like AD. Its early and accurate detection is vital for timely intervention and management, as individuals with MCI are at a higher risk of progressing to AD. However, MCI classification is particularly challenging due to its subtle and often overlapping symptoms with both CN individuals and those with AD.

In our results, the classification of MCI is particularly strong, with very few false negatives, meaning that only a small number of individuals with MCI were misclassified. This is an important achievement, as false negatives in MCI detection could lead to missed opportunities for early intervention. The high accuracy for MCI in our model demonstrates its effectiveness in correctly identifying individuals at this intermediate stage, which is essential for better monitoring and potential treatment strategies.

CNN-BiLSTM models are a type of neural network that combines the strengths of CNNs and BiLSTM networks. ViT-BiLSTM, on the other hand, is a hybrid deep learning architecture that combines the strengths of ViTs and BiLSTM networks. CNN/ViT-BiLSTM models typically work by first using a CNN/ViT network to extract spatial features from each slice of MRI. The extracted features are then fed into an LSTM network to learn the temporal dependencies between the frames.

CNNs are good at capturing local spatial patterns in images, while LSTMs are good at capturing long-term temporal dependencies in videos. ViViT models, on the other hand, are a type of transformer architecture that was specifically designed for video processing. CNNs are famous for their ability to capture local patterns and features in images via convolutional layers, whereas ViTs have demonstrated promise in capturing global context and relationships in images via self-attention mechanisms. Combining these attributes may result in enhanced performance in computer vision tasks. Also, using 3CNN can handle the long sequences of frames (slices) encountered in video (3D-MRI). This model is end-to-end in the sense that it can take raw video data as input and produce the desired output directly, without the need for any intermediate steps or preprocessing. This is in contrast to traditional video processing pipelines, which typically involve multiple separate stages, such as feature extraction, tracking, and classification.

CNN/ViT-BiLSTM models are computationally expensive to train and deploy due to the need for two separate neural networks, while ViViT models are more efficient due to their single neural network. They are also difficult to optimize due to the joint training and optimization of the two networks. However, CNN/ViT-BiLSTM models may be less accurate on some video processing tasks due to their specificity for video processing, while ViViT models are designed for general-purpose architecture.

The CNN/ViT models were trained for 100 epochs, with each individual slice undergoing 100 epochs of training. When considering the selection of 32 slices from the middle of each MRI, this amounted to a cumulative 3200 epochs for each MRI. In contrast, the ViViT model underwent a training process spanning 1500 epochs for an MRI.

ViTranZheimer outperformed the CNN-Bi-LSTM and ViT-Bi-LSTM models in AD classification accuracy, with a significant improvement of 98.6%. This improvement surpasses the CNN-Bi-LSTM model by 2.121 percentage points and the ViT-Bi-LSTM model by 1.135 percentage points, indicating a substantial enhancement in the accuracy of the ViTranZheimer model.

To ensure that the comparison of methods is reliable, it is important to use a common dataset. Most researchers choose data based on personal preference and availability. In this study, we used a standard data collection named ADNI1: Complete 3Yr 3T, which is publicly available. Using publicly available data allows other researchers to fairly compare their methods with ours using the same data collection.

## 5  Conclusions

This study introduces 'ViTranZheimer,' a novel end-to-end deep learning approach for AD classification that makes use of video vision transformers and 3D brain MRI data. To capture complex structural relationships, we model the 3D MRI volumes as videos and take advantage of the temporal dependencies between slices. Our research demonstrates the potential of deep learning models for AD diagnosis, with an emphasis on multi-class classification tasks involving NC, MCI, and AD cases. We investigated and compared various neural network architectures, such as CNN-BiLSTM, ViT-BiLSTM, and ViTranZheimer, revealing significant improvements in AD classification accuracy. The proposed approaches are thoroughly validated on ADNI dataset and compared to other relevant models. ViTranZheimer emerged as the top performer, achieving a remarkable 98.6% accuracy rate, demonstrating its diagnostic potential. The findings demonstrated that the proposed model exhibited superior performance and accuracy compared to alternative models, with a statistically significant advantage in terms of accuracy.

Future endeavors focus on MRI harmonization and ViTranZheimer to enhance AD diagnosis, enhancing accuracy and sensitivity, and fostering a comprehensive neuroimaging framework.

## Declarations

## Acknowledgments

Data collection and sharing for this project were funded by the Alzheimer's Disease Neuroimaging Initiative (ADNI), the National Institutes of Health (Grant U01 AG024904), and the DOD ADNI Department of Defense (award number W81XWH-12-2-0012). ADNI is funded by the National Institute on Aging, the National Institute of Biomedical Imaging and Bioengineering, and through generous contributions from the following: AbbVie, Alzheimer's Association; Alzheimer's Drug Discovery Foundation; Araclon Biotech; BioClinica, Inc.; Biogen; Bristol-Myers Squibb Co.; CereSpir, Inc.; Cogstate; Eisai Inc.; Elan Pharmaceuticals, Inc.; Eli Lilly and Co.; EuroImmun; F. Hoffmann-La Roche Ltd and its affiliated company Genentech, Inc.; Fujirebio; GE Healthcare; IXICO Ltd; Janssen Alzheimer Immunotherapy Research & Development, LLC; Johnson & Johnson Pharmaceutical Research & Development, LLC; Lumosity; Lundbeck; Merck & Co., Inc.; Meso Scale Diagnostics, LLC; NeuroRx Research; Neurotrack Technologies; Novartis Pharmaceuticals Corporation; Pfizer Inc.; Piramal Imaging; Servier; Takeda Pharmaceutical Company; and Transition Therapeutics. The Canadian Institutes of Health Research provides funds to support ADNI clinical sites in Canada. Private sector contributions are facilitated by the Foundation for the National Institutes of Health (http://www.fnih.org). The grantee organization is the Northern California Institute for Research and Education, and the study is coordinated by the Alzheimer's Therapeutic Research Institute at the University of Southern California. ADNI data are disseminated by the Laboratory for Neuro Imaging at the University of Southern California.

## Author Contributions Statement

MANB conceptualized and designed the study; SA and TA took part in data accumulation and data analysis; MANB, SA, TA, MSB, ED, SC, JV, and CGK wrote the manuscript; and all the authors have read, edited, and approved the article.

**Data availability:** The data are available on the ADNI website for download.

**Funding:** This work was supported by an Institutional Development Award (IDeA) from the National Institutes of General Medical Sciences NIH under grant number P20GM121307 to MANB, NIH grants R01HL172970, R01HL145753, R01HL145753-01S1, and R01HL145753-03S1 to MSB; and Institutional Development Award (IDeA) from the National Institutes of General Medical Sciences of the NIH under grant number P20GM121307 and R01HL149264 to CGK. This project is also partially supported the project Ike Muslow, MD Endowed Chair in Healthcare Informatics of LSU Health Sciences Center Shreveport.

**Ethical Approval and Consent to participate:** Not applicable.

**Human Ethics:** Not applicable.

**Consent for publication:** Not applicable.

**Competing Interests:** The authors declare no potential competing interests.

**Replication of results:** The codes and data used are available on request to enable the method proposed in the manuscript to be replicated by readers.

**Taymaz Akan** received his Ph.D. from the Department of Computer Engineering at Gazi University, Ankara, Turkey, in 2019. Currently he is a postdoctoral fellow at LSU Health Shreveport. He is experienced in medical imaging, deep learning, and optimization algorithms. His research focus is on 3D image analysis and developing disease progression models.

**Sait Alp** is an Assistant Professor in the Department of Artificial Intelligence Engineering at Trabzon University, Turkey. He earned his M.Sc. and Ph.D. degrees in Computer Engineering from Karadeniz Technical University, Trabzon, Turkey, in 2012 and 2018, respectively. His research focuses on computer vision, machine learning, biomedical image and signal processing.

**Md. Shenuarin Bhuiyan** is a professor in the department of pathology at LSU health Shreveport. His research focus is to understand the molecular and functional basis for heart performance in both health and disease.

**Elizabeth A. Disbrow** is a professor of Neurology and also the director of Center for Brain Health at LSU Health Shreveport.

**Steven A. Conrad** is a Professor of Internal Medicine at LSU health Shreveport. He is also an Ike Muslow Endowed Chair, and the Division Chief of Clinical Informatics at LSU health Shreveport.

**John A. Vanchiere** is the senior associate dean of clinical Research in the School of Medicine at LSU Health Shreveport.

**Christopher G. Kevil** is a professor in the department of pathology at LSU health Shreveport. His research focus on how the gasotransmitter hydrogen sulfide (H2S) regulates ischemic vascular remodeling and how its metabolites are affected during clinical cardiovascular disease

**Mohammad A. N. Bhuiyan** is an Assistant Professor of Internal Medicine at LSU health Shreveport. Dr. Bhuiyan's laboratory research focuses on understanding the heterogeneous effects of social determinants on cardiovascular, psychiatric, and neurobehavioral health outcomes. In addition, he is working on image optimization and image processing to develop a disease progression model for Cardiovascular (e.g., cardiomyopathy and heart failure), neurodegenerative (e.g., Alzheimer's disease), and addiction (e.g., methamphetamine and cocaine) research using machine learning and deep learning methods.